\documentclass[aps, pre, reprint]{revtex4-1}

\pdfoutput=1
\usepackage{amsmath}    
\usepackage{graphicx}   
\usepackage{verbatim}   
\usepackage{color}      
\usepackage{subfigure}  
\usepackage{hyperref}   
\begin{document}

\title{Dark solitons in laser radiation build-up dynamics}
\author{R. I. Woodward and E. J. R. Kelleher}
\affiliation{Femtosecond Optics Group, Department of Physics, Blackett Laboratory, \\ Imperial College London, London SW7 2BW, UK}

\date{\today}

\begin{abstract}
We reveal the existence of slowly-decaying dark solitons in the radiation build-up dynamics of bright pulses in all-normal dispersion mode-locked fiber lasers, numerically modeled in the framework of a generalized nonlinear Schr\"{o}dinger equation.
The evolution of noise perturbations to quasi-stationary dark solitons is examined, and the significance of background shape and soliton-soliton collisions on the eventual soliton decay is established.
We demonstrate the role of a restoring force in extending soliton interactions in conservative systems to include the effects of dissipation, as encountered in laser cavities, and generalize our observations to other nonlinear systems.
\end{abstract}

\maketitle

\section{Introduction}

Within nonlinear dispersive systems, solitons are one of the most widely studied forms of excitation: self-reinforcing solitary waves stabilized by a balance between spreading and focusing effects.  
Observations of solitons have been reported in a broad range of distinct environments including water, Bose-Einstein condensates (BECs) and optical fibers. 
Despite striking differences between these domains, a common mathematical framework described by a nonlinear Schr\"{o}dinger equation (NLSE) permits shared insight through analogies rooted in the physical description of waves~\cite{Dudley2014}.

In fiber, the solutions to the NLSE in regions of anomalous and normal group-velocity dispersion are bright and dark solitons, respectively.
While bright optical solitons are routinely observed, their dark counterparts are less well studied~\cite{Mollenauer1980,Emplit1987,Krokel1988,Weiner1988}.
True dark solitons -- satisfying the soliton condition of the defocusing NLSE -- lie on an infinite continuous-wave (CW) background.
It has been shown, however, that a dip in the intensity profile of a bright pulse is sufficient to support adiabatic dark soliton evolution, provided the width of the dip is less than a tenth of the bright pulse duration~\cite{Tomlinson1989,Kivshar1994}.

In addition to studies of temporal solitons in transmission, a wealth of nonlinear physics has been revealed through soliton dynamics in resonant cavities (e.g. mode-locked fiber lasers), where the soliton is subject to dissipative effects and other periodic perturbations.
Notably, this has included observation of rogue waves \cite{Soto-Crespo2011, Zaviyalov2012}, optical turbulence \cite{Turitsyna2013}, soliton explosions \cite{Runge2015}, stable dark pulse generation~\cite{Zhang2009c,Sylvestre2002,Ablowitz2011}, and the manifestation of self-organization effects supporting a variety of localized bright and dark soliton structures~\cite{Churkin2015}.
Analysis of such rich nonlinear phenomena is leading to breakthroughs in understanding of wave dynamics and, in the context of lasers, improved system performance.
One example of leveraging a technological benefit is the concept of long-cavity, all-normal dispersion fiber lasers for power-scaling short-pulse sources~\cite{Renninger2008a, Kobtsev2008, Kelleher2009, Erkintalo2012, Woodward_ol_2015_gco, Woodward_ptl_2014}.   
Cavity elongation, however, can lead to an increased susceptibility towards a noisy, partially mode-locked steady-state due to random dephasing of the large number of longitudinal modes~\cite{Churkin2015}.

To date, the majority of studies exploring dynamics in fiber lasers have focused on this noisy pulse state. 
Recently, however, we observed a rich dynamic -- as yet, relatively unstudied -- exists in the radiation build-up regime~\cite{Kelleher2014,Woodward_cleo15_ds}, highlighted by the emergence of quasi-stationary dark soliton structures.
Here, we numerically study the behavior and correlation of these dark structures, revealing a tendency towards self-organization and localization, and explore mechanisms for their eventual decay.
In addition to fundamental interest, greater understanding of the nonlinear physics governing this dynamic that delays the onset of steady-state laser operation could have practical implications for the control and manipulation of laser radiation build-up, and the stabilization of dark pulse sources.

\begin{figure*}[tbph]
	\centering
	\includegraphics{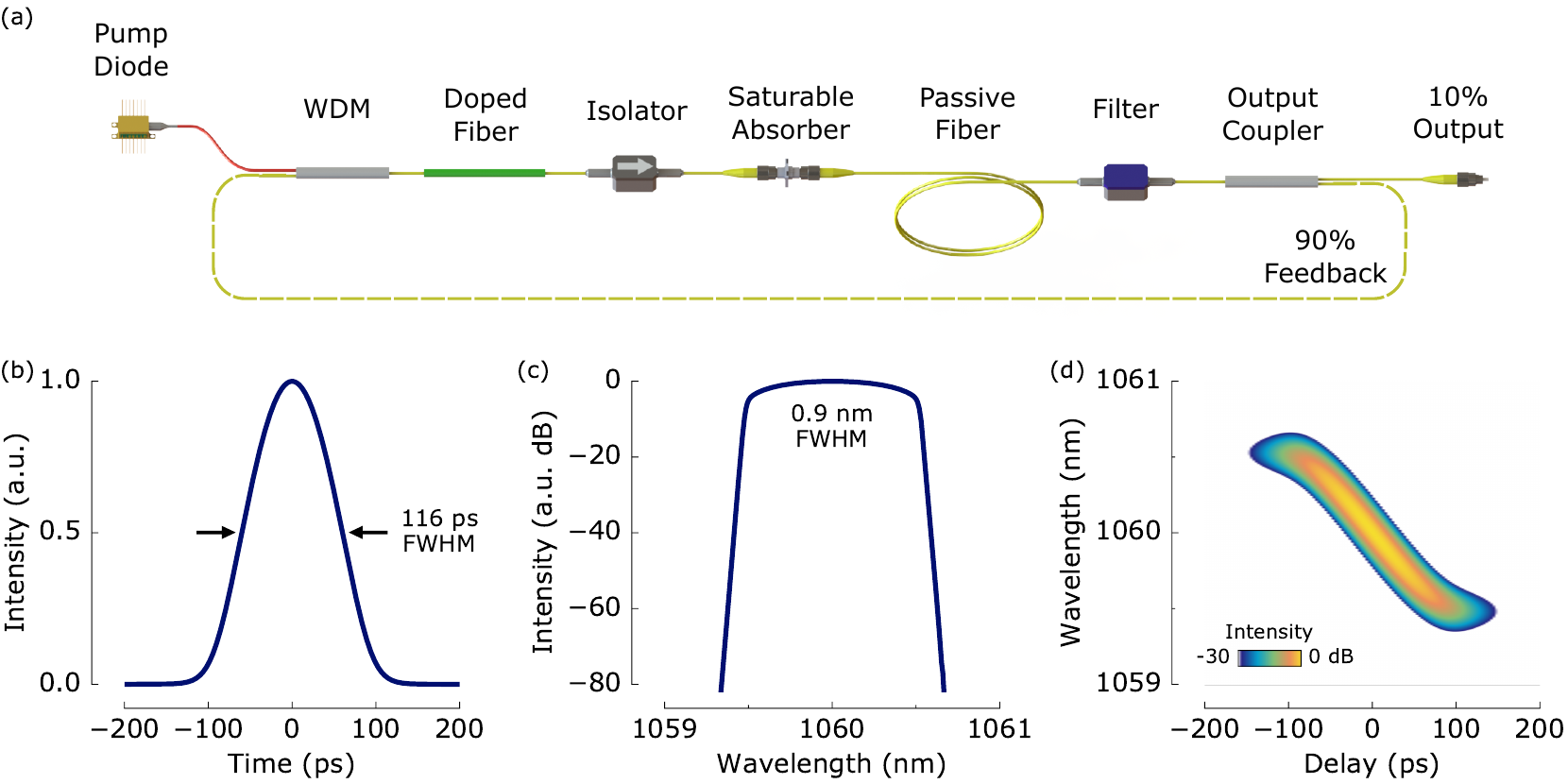}
	\caption{Long-cavity mode-locked fiber laser and steady-state properties: (a) cavity schematic; (b) pulse profile; (c) spectrum; (d) spectrogram.}
	\label{fig:cavity}
\end{figure*}

\section{Numerical Methods}

\subsection{Mode-Locked Fiber Laser Model}
We study an all-normal dispersion mode-locked fiber laser: a design which has been widely reported as a scalable high-energy pulse source.
A unidirectional ring cavity scheme is adopted, shown in Fig.~\ref{fig:cavity}(a), including a 1~m long ytterbium-doped fiber amplifier, isolator, saturable absorber, 10~nm bandpass filter centered at 1060~nm, 10\% output coupler, and a length of passive fiber.
We choose a 120~m length of passive fiber (total cavity length of 121~m) for operation in the long-cavity laser regime.

Our model propagates a discretized complex field envelope $A(z,T)$, stored on a numerical grid, sequentially through each cavity element, iterating until a steady-state regime is reached. 
The initial conditions correspond to shot noise (based on a one photon per mode model, with an associated random phase).
The field envelope in the frequency domain $A(z,\omega)$ is related to $A(z,T)$ by the Fourier transform ${\cal F}$.
Fiber sections are modeled by a generalized NLSE (Eqn.~1) formulated in the spectral domain to enable facile inclusion of frequency-dependent quantities~\cite{Travers2010}.
A scalar model is used, which assumes propagation along a principal axis of polarization-maintaining fiber.
All fibers have a propagation constant $\beta(\omega)$ and nonlinear parameter $\gamma(\omega)$, computed using a characteristic equation eigenmode analysis of a typical step-index single-mode fiber: at 1060~nm, $\beta_2=21.3$~ps$^2$ km$^{-1}$ and $\gamma=4.9$~W km$^{-1}$.
Frequency-dependent gain $g(\omega)$ in active segments is computed using a semi-analytic model based on the spectroscopic properties of a ytterbium-doped fiber amplifier and experimental pump power values~\cite{Barnard1994,Runge2014a}.
Raman scattering is included using the impulse response function $h_R$ of a multi-vibrational-mode model, with fractional Raman contribution $f_r=0.18$~\cite{Hollenbeck2002}.

We solve Eqn.~1 in a reference frame moving at the group velocity of the central grid frequency $\omega_0$, corresponding to a wavelength of 1060~nm, using an efficient embedded Runge-Kutta in the Interaction Picture scheme, with adaptive step-size control~\cite{Balac2013}.

\begin{figure*}[htpb]
	\centering
	\includegraphics{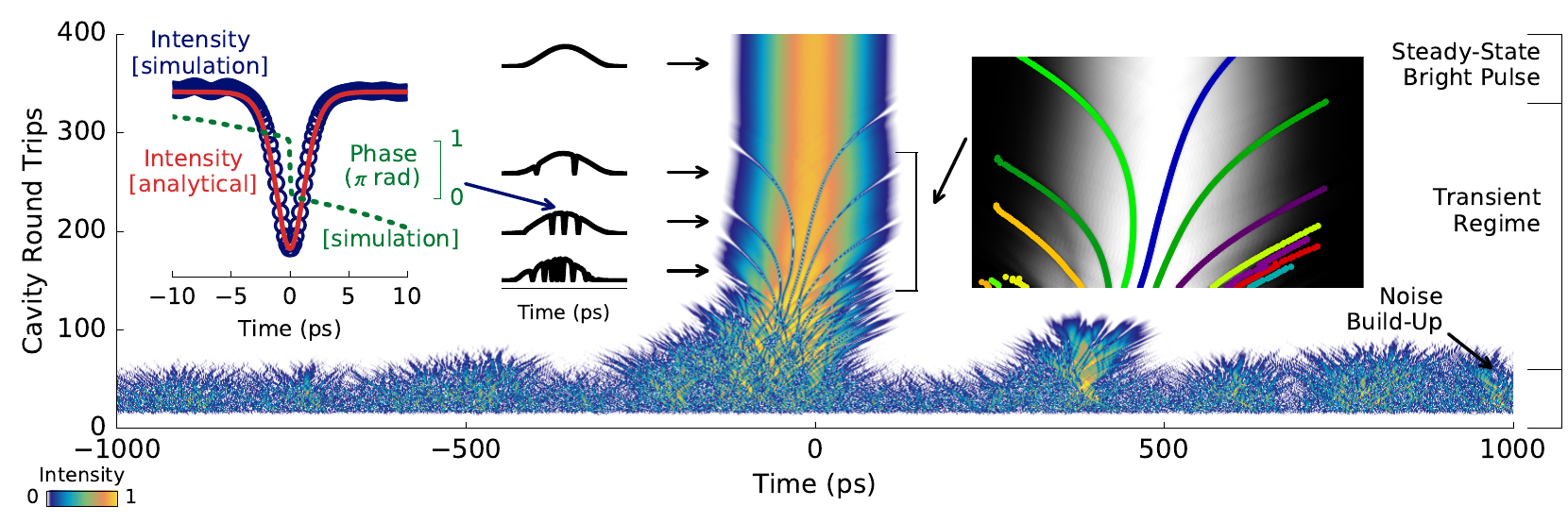}
	\caption{Temporal evolution of laser radiation build-up. Left inset: pulse profile at various iterations and magnified region showing a dark soliton fitted by its analytic expression. Right inset: magnified region of the evolution showing the tracked dark solitons overlaid as colored trajectories.}
	\label{fig:time_evolution}
\end{figure*}

\begin{widetext}
	\begin{eqnarray}
		\begin{split}
			\frac{\partial A(z,\omega)}{\partial z} =& \frac{g(\omega)}{2}A(z,\omega) + i[\beta(\omega) - \beta(\omega_0) - \beta_1(\omega_0)(\omega-\omega_0)]A(z,\omega) \\
			& + i\gamma\left[1+\frac{\omega-\omega_0}{\omega_0}\right] {\cal F}\left\{A(z,T)\left[(1-f_r)|A(z,T)|^2 + f_r\int_{-\infty}^{+\infty}h_R(T')|A(z,T-T')|^2 dT'\right]\right\}	
		\end{split}          
	\end{eqnarray}
\end{widetext}

Non-fiber components are modeled as transfer functions, including: an intensity dependent absorption operator to describe the instantaneous response of the saturable absorber, $\alpha(I)=\alpha_\mathrm{s} / (1 + I/I_\mathrm{s}) + \alpha_\mathrm{ns}$, where $I=\int|A(z,T)|^2 dT$, with modulation depth $\alpha_\mathrm{s}=15$\%, saturation intensity $I_\mathrm{s}=10$~MW cm$^{-2}$ and 10\% non-saturable loss $\alpha_\mathrm{ns}$ (chosen as typical values for state-of-the-art saturable absorbers~\cite{Woodward2015_as_2d}); a 10~nm Gaussian-shaped bandpass filter applied in the frequency domain; and implicit inclusion of optical isolation through the intrinsic unidirectionality of the model.
One round-trip of the simulated laser cavity corresponds to traversing all components in the model, after which 10\% of the field is taken as the output, with the remainder fed back for the next iteration.

Previously, we have validated this model by observing excellent agreement between simulation and experimental laser performance~\cite{Woodward_ol_2015_gco}.
For the 121~m-long cavity considered here, the modeled output properties are summarized in Fig.~\ref{fig:cavity}(b)--(d), indicating pulse generation with 116~ps full width at half maximum (FWHM) duration, 0.9~nm spectral bandwidth, and a characteristic predominantly linear chirp.

\subsection{Dark Soliton Analysis}
The temporal dynamics during radiation build-up are illustrated by recording the field envelope after each round trip (Fig.~\ref{fig:time_evolution}).
Three distinct regimes are identified: an initially disordered phase ($\sim$0--50 iterations) where the shot noise seed field is amplified; a transient regime ($\sim$50--330 iterations) as the effect of the saturable absorber and gain saturation localize the field within a highly-structured temporal envelope; and finally, a steady state ($>$330 iterations) after the internal pulse structure decays to leave a coherent, highly chirped bright pulse.

\begin{figure}[bp]
	\centering
	\sffamily
	\includegraphics{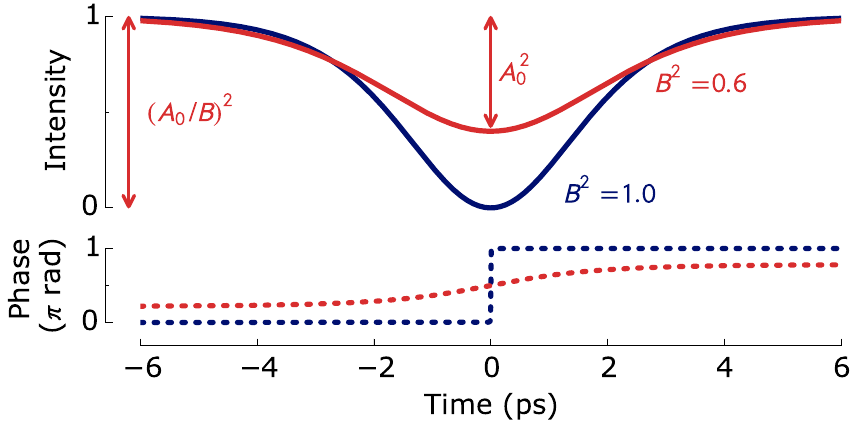}
	\normalfont
	\caption{Definition of dark soliton, showing the intensity and phase of a black ($|B|=1$) and a gray ($|B|=0.6$) soliton. The notation of Eqn.~\ref{eqn:ds} is illustrated by arrows, which relate to the gray soliton.}
	\label{fig:ds_analytical}
\end{figure}

In the transient regime we observe solitary dark structures as holes in the background intensity, co-existing in the early evolution with stochastic radiation and remarkably, persisting for hundreds of round trips as quasi-stationary structures before decaying.
We discover that these structures are dark solitons by observing strong agreement between their amplitude and phase profile and the theoretical definition of a dark soliton, given by~\cite{Tomlinson1989}:
\begin{equation}
\label{eqn:ds}
A(T) = A_0 \sqrt{B^{-2} - \mathrm{sech}(T/T_0)^2} \times \exp[i \phi(T/T_0)]
\end{equation}
where $A_0^2$ is the intensity dip and $B$ is the blackness parameter ($0\leq |B| \leq 1$) defining the ratio of the dip minimum to the background, such that the background power level (measured in watts) is $(A_0/B)^2$, as illustrated in Fig.~\ref{fig:ds_analytical}. 
The $1/e$ soliton duration is related to the fiber parameters by $T_0 = \sqrt{\beta_2/(A_0^2 \gamma)}$ and the phase is:
\begin{equation}
\phi(T/T_0) = \sin^{-1}\left(\frac{B \mathrm{tanh}(T/T_0)}{\sqrt{1 - B^2 \mathrm{sech}^2(T/T_0)}}\right)
\end{equation}
Fig.~\ref{fig:time_evolution} left inset shows that a dark structure during radiation build-up is well-fitted by the shape of a black ($|B|=1$) soliton, in addition to possessing the expected anti-symmetric $\pi$ phase shift at its center~\cite{Kivshar1998}.
We confirm that all persistent dips in intensity fit the functional form of a black ($|B|=1$) or gray ($|B|<1$) soliton.

To probe this phenomena, an ensemble of 600 independent realizations of the same simulation is executed, each starting from a different randomized shot noise field.
All realizations converge to the same steady-state pulse, but the build-up dynamics and the number of round trips required for the pulse to display negligible iteration-to-iteration changes are found to vary. 
The laser cavity configuration and component parameters establish the basin of attraction for the system, creating a fixed-point attractor (in our case, a stable linearly-chirped bright pulse) towards which initial conditions converge.
Thus, in order to study the persistence of localized patterns in the transient regime we develop an analysis technique for isolating and tracking the dark solitons.

Briefly, the algorithm searches all iterations of every simulation to identify possible dark soliton structures by their shape; this is proceeded by a sorting process to group structures that, on consecutive round-trips, appear within the same bounded location of time, and can thus be considered consistent.  
If a candidate dark structure cannot be found in the future iteration it is deemed to have disappeared (i.e. decayed); similarly, if a structure is identified with no correlation to the previous iteration it is assigned a unique identity, and can be considered as the birth of a new dark soliton.
This process repeats until all candidate dark structures of a two-dimensional map are assigned to uniquely identifiable dark solitons, enabling their motion through the bright pulse envelope to be tracked.
Fig. \ref{fig:time_evolution} right inset shows a section of a processed evolution, with colored dots indicating the trajectories of unique dark solitons; visual inspection confirms that the algorithm correctly identifies and tracks the dark structures.

\section{Dark Soliton Creation and Decay}
Although unexpected, it is not surprising \textit{a posteriori} that dark solitons are observed: spontaneous noise perturbations create intensity dips, which evolve into dark solitons in the presence of normal dispersion since these are the stable solutions to the defocusing NLSE~\cite{Hasegawa1973b}. 
This is analogous to arbitrary bright waveforms converging towards soliton solutions in the anomalous dispersion regime~\cite{Gouveia-Neto1989}, although dark soliton generation is thresholdless and thus occurs more readily~\cite{Gredeskul1990}.

In all simulations, the dark solitons eventually decay; thus the system steady-state is a coherent bright pulse.
Decay is interpreted as the walk-off of a dark soliton from the bright pulse, arising from a mismatch in the group velocity of the soliton relative to its background.
This behavior is expected for lighter gray ($|B|<1$) solitons, which have a shallower and more gradual transition in their temporal phase, and hence a greater soliton-background group velocity mismatch~\cite{Kivshar1998}.
In our simulations, however, we identify that the majority of dark structures that form are in fact dark gray or black solitons, possessing a group velocity equal or close to the background, and should thus propagate stably.
How can we explain their decay?

Stimulated Raman scattering (SRS) is well known to contribute to the decay of dark solitons~\cite{Kivshar1991,Horikis2015a}; however, here we observe soliton decay even in the absence of SRS.
Therefore, we propose an explanation by considering the shape of their background (i.e. the bright pulse) and note that a sloped background intensity across a dark soliton will impart an additional phase change, resulting in a gradient-dependent group velocity relative to the velocity of its background~\cite{Kivshar1994}.
Due to the spontaneous noise-seeded origin of the dark solitons in our system, they are distributed throughout the bright pulse.
Consequently, the majority exist in the pulse wings, embedded in a non-uniform background such that the intensity dips are accelerated by the background gradient.
This acceleration is indicated by curved trajectories of the decaying solitons as they walk off from the bright pulse (Fig.~\ref{fig:time_evolution}).
Finally we note that, although the peak of the bright pulse can be considered a point of unstable equilibrium, where a black soliton experiences a near-uniform intensity and consequently has a negligible soliton-background group velocity mismatch, perturbations give rise to a timing jitter, moving the soliton away from this point of equilibrium, triggering its eventual decay. 

\begin{figure}[tbp]
	\centering
	\sffamily
	\includegraphics{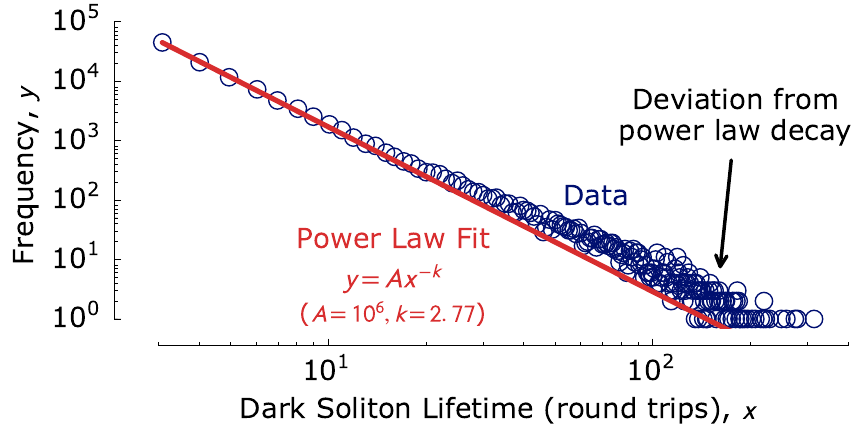}
	\normalfont
	\caption{Histogram of soliton lifetimes, showing decreasing occurrence of long-lived dark solitons (and a small number of very long-lived ones, which deviate from a power law dependence).}
	\label{fig:histogram1}
\end{figure}

It is remarkable, however, that in the presence of dissipative processes such as periodic gain and loss and subject to high-order dispersive and nonlinear perturbations (including Raman and optical shock formation), dark solitons can remain stable over many hundreds of round trips.
This corresponds to an effective propagation distance of tens of kilometers in a non-conservative soliton system, as described by a generalized NLSE. 
Using the definition of nonlinear length $L_\mathrm{NL}=(\gamma P_\mathrm{pk})^{-1}$, with peak power $P_\mathrm{pk}=1.25~W$, we find that $L_\mathrm{NL}=163$~m, confirming that these dark structures persist for hundreds of nonlinear lengths, suggesting quasi-stationarity from self-organization effects~\cite{Turitsyna2013}.

To understand the statistical nature of this phenomena, we analyze the aggregate decay rates for tracked solitons from the ensemble, where dark soliton lifetime is defined as the number of round trips between spontaneously forming and accumulating a sufficient time-delay to move outside the bright-pulse background.
The resulting histogram (Fig.~\ref{fig:histogram1}) shows that short-lived dark solitons occur much more frequently than persistent solitons with a long lifetime, and a heavy-tail distribution is observed.
We fit the empirical data with a power law, which is widely used to describe decay in complex systems, in addition to appearing in a broad range of noise-seeded physical phenomena~\cite{Clauset2009}.
The frequency $y$ is, therefore, proposed to vary with soliton lifetime $x$ as follows: $y=A x^{-k}$, with normalization factor $A=1\times10^6$ and exponent $k=2.77$.

\begin{figure}[tbp]
	\centering
	\sffamily
	\includegraphics{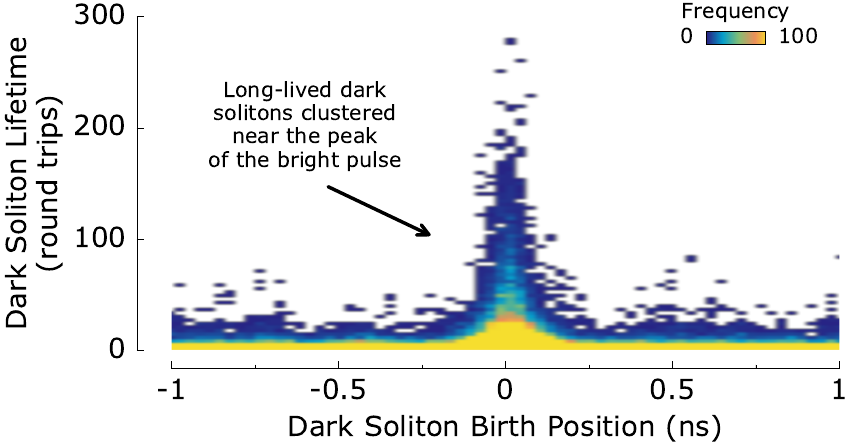}
	\normalfont
	\caption{Two-dimensional histogram of soliton lifetimes and birth positions in the bright pulse background (0 ns refers to the center of the bright pulse).}
	\label{fig:histogram2}
\end{figure}

For lifetimes shorter than approximately 50 round trips, the power law accurately describes the distribution.
Significant deviation, however, is observed for longer lifetimes.
This suggests that such persistent quasi-stationary solitons can be considered as rare events: phenomena which occur with an unlikely, but non-negligible, probability.
Rare events are particularly interesting in nonlinear systems tending towards an attractor, helping to reveal aspects of the underlying dynamics, which could ultimately be controlled for practical exploitation~\cite{Dudley2014}. 
These observations pave the way for further work to quantitatively relate the soliton decay mechanisms to their lifetime and more generally, to contribute to understanding the underlying factors responsible for rare phenomena.

To phenomenologically relate dark soliton lifetime to the proposed routes for soliton decay, we plot a 2D histogram of lifetimes relative to birth position (Fig.~\ref{fig:histogram2}). 
The birth position is defined as the temporal location where the dark soliton spontaneously forms relative to the center of the bright pulse background. 
Longer-lived solitons are seen to cluster close to the bright pulse center -- the point of unstable equilibrium in our dynamical system. 
This agrees well with our explanation of decay rate: dark solitons that form near the bright pulse peak experience a near-uniform background intensity and have a small soliton-background group velocity mismatch, yielding quasi-stationary solitons with a slow decay rate and a long lifetime.

\section{Interactions and Collisions}
\begin{figure}[htbp]
	\centering
	\includegraphics{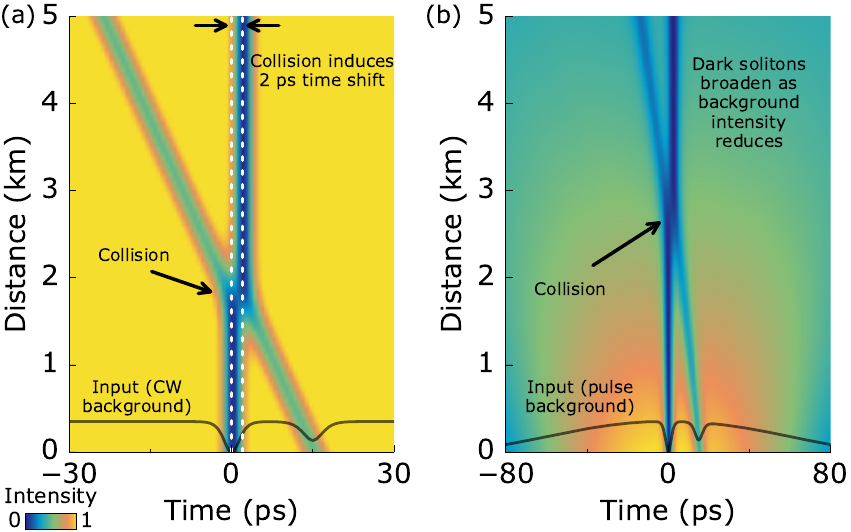}
	\caption{Collision dynamics between a black $|B|=1$ and gray $|B|=0.6$ soliton in transmission: (a) on a CW background (white dotted lines indicate the collision-induced temporal displacement of the black soliton); (b) on a bright pulse background.}
	\label{fig:compare_collisions1}
\end{figure}

\begin{figure}[tbp]
	\centering
	\includegraphics{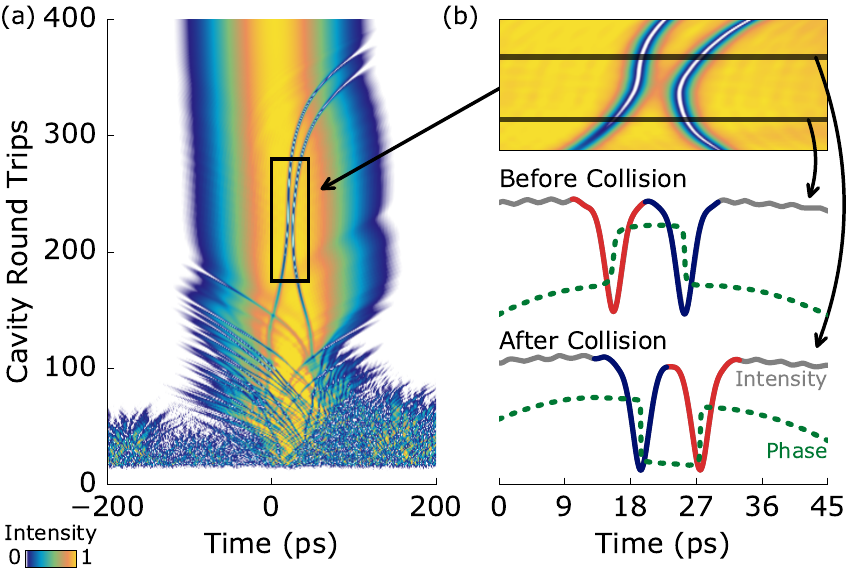}
	\caption{Radiation build-up dynamics (using the parameters as in Fig.~\ref{fig:time_evolution}, but initiated from a different randomized shot noise field): (a) temporal evolution; (b) magnified collision region and dark soliton profiles within the bright pulse, before and after a collision at 200 and 240 round trips, respectively (color is used to identify each soliton).}
	\label{fig:collisions}
\end{figure}

A defining feature of soliton solutions of the NLSE is their ability to pass through each other without changing velocity or shape as a result of collisions~\cite{Hasegawa1973b}.
The collision of ideal dark solitons (on an infinite CW background) during transmission in fiber has been shown to cause a mutual time shift, but otherwise leave their propagation unchanged~\cite{Gordon1983,Thurston1991}.
A larger time shift is observed for colliding solitons of greater blackness. 
This behavior has also been verified for dark pulses on a finite bright pulse background that satisfy the condition for adiabatic soliton propagation~\cite{Thurston1991}.

We illustrate this interaction with a black $|B|=1$ and gray $|B|=0.6$ soliton, initially separated by 15~ps, propagating along a length of passive fiber.
For the case of a CW background, the black soliton travels at the group velocity of the background (here, equal to the window of the simulation frame) while the gray soliton moves more quickly, walking into the path of the black soliton. 
It can be seen in Fig.~\ref{fig:compare_collisions1}(a) that the collision leads to a displacement of the solitons by $\sim$2~ps, relative to their initial trajectories (denoted by white dotted lines). 
A similar evolution is observed for the same soliton pair on a finite background [Fig.~\ref{fig:compare_collisions1}(b)]; however, in this case the sloped background initially slows the gray soliton, delaying the point of collision. 
It should also be noted that the bright pulse disperses leading to a monotonic reduction in the background intensity, and hence a continuous broadening of the solitons duration (as expected from the soliton condition).

\begin{figure}[htbp]
	\centering
	\includegraphics{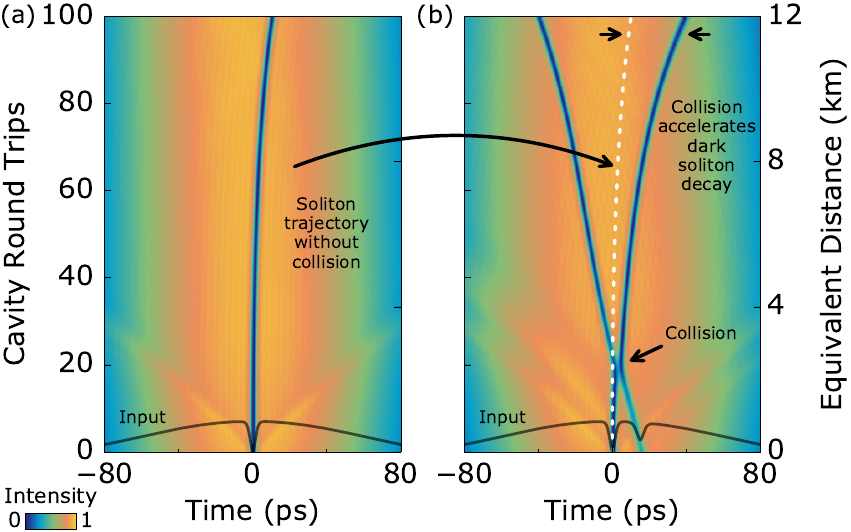}
	\caption{Dark soliton dynamics in a dissipative cavity: (a) black $|B|=1$ soliton propagation; (b) black and gray $|B|=0.6$ soliton propagation, resulting in a soliton-soliton collision. The mutual time shift induced by the collision accelerates the black solitons decay (its original trajectory highlighted by a white dotted line). The weak perturbations to the background in the early evolution are attributed to a readjustment of the steady-state cavity dynamic in the presence of a small, but non-negligible loss of energy due to the inclusion of the soliton pulses.}
	\label{fig:compare_collisions2}
\end{figure}

Interestingly, we also observe these dynamics in the radiation build-up regime of our laser -- a dissipative system described by a generalized NLSE -- where the output field extracted once per round trip represents a long-range propagation, sampled every 121~m.
Here, the emergence of dark solitons with varying blackness and position from initially random fluctuations in the laser intensity leads to a large number of collisions in the early build-up (see Fig.~\ref{fig:time_evolution}).
The noise-seeded process results in a unique evolution, even for successive turn-on instances of the same laser [e.g. Fig.~\ref{fig:collisions}(a)].
In the transient regime, isolated collision events can be identified: in Fig.~\ref{fig:collisions} after $\sim$220 round trips, two black solitons collide, move through each other while experiencing a time shift, but preserve their phase [Fig.~\ref{fig:collisions}(b)]. 
While it may appear from the intensity alone that this interaction between like solitons is repulsive, they are in fact transmitted, experiencing only temporal displacement as a result of the interaction.
The apparent node between the two solitons, resulting in a negligible intensity dip at the center of the collision, is caused by interference between the two fields~\cite{Thurston1991,Nguyen2014}.
The collision-induced time shifting of solitons is a complementary mechanism that can trigger dark soliton decay from a quasi-stationary state atop a bright pulse in a near uniform background. 

Additionally, these observations suggest that the established behavior of interacting dark solitons in conservative systems can extend to dissipative systems exhibiting gain and loss, and in the presence of higher-order nonlinearities, e.g. Raman scattering. 
We verify this by injecting the steady-state bright pulse back into the laser cavity, recirculating the field over 100 round trips, and considering two distinct cases: first, we add a black $|B|=1$ soliton and observe its evolution; second, we add a black $|B|=1$ and gray $|B|=0.6$ soliton to create a collision event. In both cases, the quasi-stationary solitons are susceptible to perturbations, mediating their eventual decay; however, the collision induced time-shift with the gray soliton [after $\sim$20 round trips in Fig.~\ref{fig:collisions}(b)] can be seen to accelerate the decay process: moving the black soliton away from the point of unstable equilibrium, and onto the slope of the background pulse.

\section{Discussion}

The recirculation of light in fiber resonators offers a unique opportunity for exploring long-range soliton interactions by periodically sampling the optical field once per round-trip~\cite{Jang2013}.
In particular, mode-locked lasers -- where a near constant bright pulse background is maintained by the restoring forces of the system -- offer a suitable environment for studying dark soliton evolution, and their interactions. 
The complexity of many-moded, long cavity lasers has now been shown to support the spontaneous emergence of quasi-stationary dark solitons that could be studied on laboratory time-scales, albeit requiring state-of-the-art techniques~\cite{Churkin2015}. 

While our analysis of dark soliton dynamics in the laser radiation build-up regime is performed on a 121~m mode-locked system, we emphasize that the conclusions are generally applicable to this type of normally dispersive laser. 
We confirm this by observing similar patterns in the evolution dynamics of cavities with 10~m to kilometer length-scales.
The longer cavities, however, which support broader bright pulses in the steady-state due to greater round trip dispersion, act to sustain the quasi-stable formation of dark solitons, increasing their discernibility. 

The NLSE underpins numerous nonlinear dispersive systems: the connection between the dynamics observed in fiber-optics and BECs is an example of one area receiving particular attention~\cite{Proukakis2004}.
Dark soliton matter-waves have been observed experimentally in BECs, where they can be a useful diagnostic for probing mesoscopic physics in ultracold gases~\cite{Frantzeskakis2010}; however, they have also been shown to be thermodynamically unstable, resulting in their eventual decay~\cite{Becker2008,Frantzeskakis2010}. 
Additionally, soliton-soliton collisions have been observed to contribute to the decay mechanism~\cite{Becker2008,Huang2001}, in analogy with the dynamics during laser radiation build-up we report here. 
We thus suggest that mode-locked fiber lasers could prove a useful platform for improving understanding of dark soliton interactions, with applicability to other dissipative systems.

In conclusion, we have observed the spontaneous creation of slowly-decaying dark solitons from noise within the radiation build-up dynamics of bright pulses in all-normal dispersion mode-locked fiber lasers.
Mechanisms have been proposed for dark soliton decay, considering the role of the background gradient and soliton-soliton interactions.
The restoring forces of mode-locked laser cavities were shown to sustain a bright pulse background, offering a platform for probing soliton collisions over long distances.	
We therefore envisage future studies using mode-locked fiber resonators to further study dark soliton dynamics, offering insight into the underlying nonlinear physics of solitons in dissipative systems.

\acknowledgements{We thank Roy Taylor for stimulating discussions. RIW acknowledges support through an EPSRC Doctoral Prize Fellowship and EJRK is supported by a Royal Academy of Engineering Fellowship.}


\end{document}